\documentclass[aps,prd,groupedaddress,twocolumn,showpacs]{revtex4-1}

\usepackage{hyperref}
\usepackage{amsfonts}
\usepackage{amsmath}
\usepackage{amssymb}
\usepackage{latexsym}
\usepackage{graphicx}
\usepackage{color}

\usepackage{soul}

\newcommand{\be}{\begin{equation}}
\newcommand{\ee}{\end{equation}}
\newcommand{\bea}{\begin{eqnarray}}
\newcommand{\eea}{\end{eqnarray}}

\newcommand{\mralpha}{{{}_{\text{o}}\alpha}}
\newcommand{\mrbeta}{{{}_{\text{o}}\beta}}

\newcommand{\Nboost}{{\mathcal{N}}_{\text{1}}}
\newcommand{\Noneway}{{\mathcal{N}}_{\text{2}}}
\newcommand{\Ntwoway}{{\mathcal{N}}_{\text{3}}}

\newcommand{\Nkick}{{\mathcal{N}}_{\text{kick}}}

\DeclareMathOperator{\atanh}{atanh}

\DeclareMathOperator{\Realpart}{Re}

\DeclareMathOperator{\polylog}{Li}

\DeclareMathOperator{\diag}{diag}

\date{May 2011, revised February 2012. 
Published in Phys.\ Rev.\ D {\bf 85}, 061701(R) (2012).\\
XX This arXiv posting corrects a typographic error in Eq.~(7a) of the published version.}

\begin{document}
\title{Voyage to Alpha Centauri: 
Entanglement degradation of cavity modes due to motion}
\author{David Edward Bruschi}
\author{Ivette Fuentes}
\thanks{Previously known as Fuentes-Guridi and Fuentes-Schuller.}
\author{Jorma Louko}
\affiliation{School of Mathematical Sciences, 
University of Nottingham, 
Nottingham NG7 2RD, 
United Kingdom}

\begin{abstract}
We propose a scheme to investigate whether non-uniform
motion degrades entanglement of a relativistic quantum field
that is localised both in space and in time. For a Dirichlet
scalar field in a cavity in Minkowski space, in small but
freely-adjustable acceleration of finite but arbitrarily
long duration, degradation of observable magnitude occurs
for massless transverse quanta of optical wavelength at
Earth gravity acceleration and for kaon mass quanta already
at microgravity acceleration. We outline a space-based
experiment for observing the effect and its gravitational
counterpart.
\end{abstract}


\maketitle

\section{Introduction} 

A common way to implement quantum
information tasks involves storing information in cavity field
modes. How the motion of the cavities affects the stored information
is a question that could be of practical relevance in space-based
experiments~\cite{Villoresi:2008p491,*Bonato:2009p959}.

In this paper we analyse the degradation of initially maximal
entanglement between scalar field modes in two Dirichlet cavities in
Minkowski space, one inertial and the other undergoing motion that
need not be stationary. Our analysis combines, to our knowledge for
the first time in relativistic quantum information theory, the
explicit confinement of a quantum field to a finite size cavity and a
freely adjustable time-dependence of the cavity's acceleration. This
allows observers within the cavities to implement quantum information
protocols in a way that is localised both in space and in
time~\cite{Downes:2010tv}. In particular, our system-environment split
is manifestly causal and invokes no horizons or other teleological
notions that would assume acceleration to persist into the asymptotic,
post-measurement future
(cf.~\cite{Schlicht:2003iy,*Langlois:2005nf,*Louko:2006zv,*Satz:2006kb,*Louko:2007mu}).
By the equivalence principle, the analysis can be regarded as a model
of gravity effects on entanglement.

Our main results are:

\begin{itemize}
\item[(i)]
In generic motion, particle creation in the moving cavity causes
the entanglement to depend on time.  This is in stark contrast to the
previously-analysed relativistic situations 
(see~\cite{Alsingtelep,FuentesSchuller:2005p47,*PhysRevA.82.042332} 
for a small sample) where uniform acceleration is assumed to persist
into the asymptotic future and the entanglement between inertial and
accelerated observers is argued to be preserved in time.
\item[(ii)]
We give an analytic method for computing the entanglement for
trajectories that consist of inertial and uniformly linearly
accelerated segments in the small acceleration approximation. 
An advantage over the small amplitude approximations 
customary with the dynamical Casimir effect 
\cite{PhysRevLett.77.615,*wilson-casimir-obs,*dodonovCasimirReview2010}
is that the segment durations 
and the travelled distances are arbitrary. 
The entanglement remains constant within each segment,
but we find that it does depend on the \emph{changes\/} 
in the acceleration and on the time intervals between these changes. 
We present explicit results for three
sample scenarios for a massless field in $(1+1)$ dimensions, finding
in particular that in this approximation any degradation caused by the
accelerated segments can be undone by fine-tuning the durations of the
intermediate inertial segments.
\item[(iii)]
In $(3+1)$ dimensions, the entanglement degradation has an
observable magnitude for massless transverse quanta of optical
wavelength at Earth gravity acceleration and for kaon mass quanta
already at microgravity acceleration.  The effect should hence be
detectable in space-based experiments, where it would in particular
test whether an equivalence principle between acceleration and a
gravitational field holds also in the context of quantum information.
\end{itemize}

\section{Cavity prototype configuration} 

Let Alice and Rob 
(``Relativistic Bob''~\cite{Alsingtelep}) 
be observers in $(1+1)$-dimensional Minkowski
spacetime, each carrying a cavity that contains an uncharged
scalar field of mass $\mu\ge0$ 
with Dirichlet boundary conditions. Alice and Rob are
initially inertial with vanishing relative velocity, and each cavity
has length $\delta>0$. The field modes in each cavity are discrete,
indexed by the quantum number $n\in \{1,2,\ldots\}$ and having the
frequencies $\omega_n := \sqrt{M^2 + \pi^2 n^2} / \delta$ 
where $M := \mu\delta$ 
(we set $c = \hbar = 1$).

Let Alice and Rob initially 
prepare their two-cavity system in the 
maximally entangled Bell state
$\left|\Psi\right>=\tfrac{1}{\sqrt{2}} 
\bigl(\left|0\right>_{A}\left|0\right>_{R}
+\left|1_{k}\right>_{A}\left|1_k\right>_{R}\bigr)$, 
where the subscripts $A$ and $R$ identify the cavity, 
$\left|0\right>$~is the vacuum and 
$\left|1_k\right>$ is the one-particle state with
quantum number~$k$. Experimentally, $\left|\Psi\right>$ might be
prepared by allowing a single atom to emit an excitation of frequency
$\omega_k$ over a flight through the two
cavities~\cite{Raimond:2001p6,*Browne:2003p374}, 
and the assumption of
a single $k$ 
is experimentally justified if $\delta$ is so small that
cavity's frequency separation $\omega_{n+1} - \omega_{n}$ 
is large compared with the frequency separations of the atom.

\begin{figure}
\includegraphics[scale=0.8]{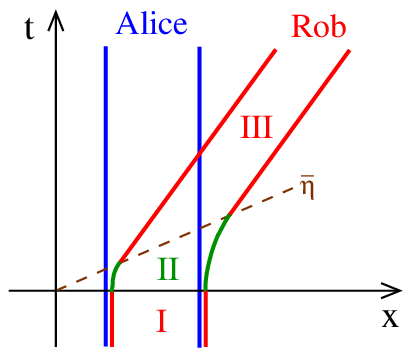}
\caption{\label{PIchi}Cavity trajectories in Minkowski space. 
Alice's cavity remains inertial. 
Rob's cavity is inertial in region~I, accelerates in region II
and is again inertial in region~III\null. 
The figure shows the prototype case where 
Rob's acceleration is to the right and uniform throughout 
region~II, and $\bar\eta$ is the duration of the 
acceleration in Rindler time $\atanh(t/x)$.}
\end{figure}

We then allow Rob to undergo arbitrary acceleration for a finite
interval of his proper time. After the acceleration Rob's cavity is
again inertial and has proper length $\delta$ in its new rest frame. 
Figure \ref{PIchi} shows the prototype case where Rob's acceleration is
uniform while it lasts.

Let $U_n$, $n=1,2,\ldots$, denote Rob's field modes that are of
positive frequency $\omega_n$ with respect to his proper time before
the acceleration. Let $\bar{U}_n$, $n=1,2,\ldots$, similarly
denote Rob's field modes that are of positive frequency $\omega_n$
with respect to his proper time after the acceleration. 
The two sets of modes are related by the Bogoliubov transformation 
$\bar{U}_m = \sum_n \, \bigl( 
\alpha_{mn} U_n + \beta_{mn} U^*_n
\bigr)$, 
where the star denotes complex conjugation and the 
coefficient matrices $\alpha$ and $\beta$ are determined by the motion
of the cavity during the acceleration~\cite{birrell-davies}. 
Working in the Heisenberg picture, the state $\left|\Psi\right>$ does
not change in time, but for late time observations the early time
states $\left|0\right>_{R}$ and $\left|1_k\right>_{R}$ need to be
expressed in terms of Rob's late time vacuum
$\left|\bar{0}\right>_{R}$ and the late time excitations on it, by
formulas that involve the Bogoliubov
coefficients~\cite{fabbri-navarro-salas}. In this sense, the
acceleration creates particles in Rob's cavity.

We regard the late time system as tripartite between Alice's cavity,
the (late time) frequency $\omega_k$ in Rob's cavity and the (late time)
frequencies $\{\omega_n \mid n\ne k\}$ in Rob's cavity.  As any
excitations in the $n\ne k$ frequencies are entirely due to the 
acceleration, we regard these frequencies as the environment. 
We ask: Has the entanglement between Alice's cavity
and the frequency $\omega_k$ in Rob's cavity been degraded, 
from the maximal value it had before Rob's acceleration?

We quantify the entanglement by the negativity, the widely-used 
entanglement monotone defined by 
$\mathcal{N}(\rho)
:=-{\textstyle \sum_{\lambda_\text{I}<0}} \, \lambda_\text{I}$, where
the reduced density matrix $\rho$ is obtained by tracing the full
density matrix $\left|\Psi\right> \left< \Psi \right|$ over Rob's late
time frequencies $\{\omega_n \mid n\ne k\}$ and $\lambda_\text{I}$ are
the eigenvalues of the partial transpose of~$\rho$. $\mathcal{N}$~has
the advantage of being easy to compute in bipartite systems of
arbitrary dimension~\cite{vidal-werner:computable}. The
closely-related logarithmic negativity, $E_{\mathcal{N}} :=
\ln\bigl(1+\mathcal{N}\bigr)$, is an upper bound on the distillable
entanglement $E_D$ and is operationally interpreted as the
entanglement cost $E_C$ under operations preserving the positivity of
partial transpose~\cite{audenart-plenio-eisert:cost}. In this respect,
the entanglement quantification based on negativity nicely
interpolates between the two canonical (yet generically difficult to
compute) extremal entanglement measures $E_D$ and
$E_C$~\cite{plenio-virmani:review}.  A~sample of recent entanglement
analyses utilising negativity can be found
in~\cite{PhysRevLett.106.040502,*PhysRevLett.106.040503,%
*PhysRevLett.106.080502,*PhysRevLett.106.110402,*PhysRevLett.106.120404}.

The situation covering all the scenarios below 
is when 
$\alpha = \diag(z_1,z_2,\ldots) + h\alpha^{(1)} + O(h^2)$
and 
$\beta = h\beta^{(1)} + O(h^2)$, 
where $h$ is a small parameter, 
the first-order coefficient matrices 
$\alpha^{(1)}$ and $\beta^{(1)}$ are off-diagonal, and each 
$z_n$ has unit modulus. 
We find that the first correction to $\rho$ occurs in order~$h^2$, 
the partial transpose is to this order an 
$8\times8$ matrix with exactly one negative eigenvalue, and the 
order $h^2$ formula for the negativity reads 
\begin{align}
\textstyle 
\mathcal{N}
= \tfrac12 
- h^2 
\sideset{}{'}\sum_n 
\left( 
\frac12 
\bigl|\alpha^{(1)}_{nk}\bigr|^2 
+ \bigl|\beta^{(1)}_{nk}\bigr|^2 
\right) , 
\label{Roku}
\end{align}
where the prime on the sum means that the 
term $n=k$ is omitted.

\section{Massless field} 

We now specialise to a massless field. 
Let I, II and III
denote respectively the initial inertial region, the region of
acceleration and the final inertial region. 
As a first travel scenario, let the acceleration
in region II be uniform, 
in the sense that Rob's cavity is dragged to
the right by
the boost Killing vector 
$\xi := x\partial_t + t\partial_x$ as shown in
Figure~\ref{PIchi}. Let the proper acceleration at the centre of the
cavity be~$h/\delta$, where the dimensionless positive parameter $h$
must satisfy 
$h<2$ in order that the proper acceleration 
at the left end of the cavity is finite. In region~II, the 
field modes that are positive frequency with respect to 
$\xi$ are then a discrete set $V_n$ with 
$n\in \{1,2,\ldots\}$
and their frequencies with respect to the proper time $\tau$ 
at the centre of the cavity are 
$\tilde\Omega_n = (\pi h n)/[2\delta\atanh(h/2)]$. 

The Bogoliubov transformation from 
$\{U_n\}$ to 
$\{V_n\}$ can be computed by standard techniques \cite{birrell-davies} 
at the junction $t=0$. The coefficient matrices, which we denote by 
$\mralpha$ and~$\mrbeta$, have small $h$ expansions that begin 
\begin{subequations}
\label{eq:mrcoeffs}
\begin{align}
\mralpha_{nn} 
&= 
1-{\tfrac {1}{240}}\,{\pi }^{2}{n}^{2}{h}^{2}
+ O(h^4) , 
\label{ShiNi}
\\[1ex]
\mralpha_{mn} 
&= 
\sqrt {mn} \, 
{\frac { \bigl( -1+ {( -1 )}^{m-n}\bigr) }{{\pi }^{2} {( m-n )}^{3}}}h
+ O(h^2) 
\hspace{2ex}(m\ne n) , 
\label{ShiIchi}
\\[-1ex]
\mrbeta_{mn} 
&= 
\sqrt {mn} \, 
{\frac { \bigl(1 - {( -1 )}^{m-n} \bigr)}{{\pi}^{2} 
{( m+n )}^{3}}}h
+ O(h^2) . 
\label{ShiSan}
\end{align} 
\end{subequations}

The Bogoliubov transformation from region I to region 
III can now be written as the composition of 
three individual transformations. 
The first is with the coefficient matrices 
$(\mralpha,\mrbeta)$ from I to~II\null. 
The second is with the coefficient matrices
$\bigl(\diag(p, p^{2}, p^{3},\ldots),\diag(p^{-1}, p^{-2},
p^{-3},\ldots)\bigr)$, where 
$p:=
\exp\bigl(i\tilde\Omega_1 \bar{\tau}\bigr)$ 
and $\bar\tau$ is the duration of
the acceleration in~$\tau$: this undoes the phases that
the modes $V_n$ develop over region~II\null. 
The third is the inverse of the first, 
from II to~III,  
with the coefficient matrices 
$\bigl(\mralpha^\dagger,-\mrbeta^T\bigr)$.  
Collecting, we find 
from (\ref{Roku})  
that the negativity $\Nboost$ 
is given to order $h^2$ by 
\begin{align}
\Nboost
&= \tfrac12 
- h^2 \sum_{r=0}^{\infty}
a_{kr} \, {\bigl|p^{1+2r} - 1\bigr|}^2  
\notag
\\[1ex]
&= \tfrac12 
- 2 \bigl[Q(k,1) - Q(k,p) \bigr] h^2 , 
\label{Hachi}
\end{align}
where 
\begin{align}
& Q(n,z) := \frac{4n^2}{\pi^4}
\Realpart\left(\polylog_6(z) - \frac{1}{64}\polylog_6(z^2)\right)
\notag
\\[1ex]
& \ 
+ \frac{6n}{\pi^4}\sum_{r= 
\left[
\frac{n}{2}
\right]}^{\infty}
\Realpart\left(z^{1+2r}\right)\left(\frac{1}{{(1+2r)}^5} 
- \frac{n}{{(1+2r)}^6} \right) 
\label{Ku}
\end{align}
and $a_{nr}$ 
are the coefficients in the expansion 
$Q(n,z) = \sum_{r=0}^{\infty}
a_{nr}\Realpart\left(z^{1+2r}\right)$. 
Here $\polylog_6$ is the polylogarithm~\cite{nist-dig-library}, 
the square brackets in the lower limit of the sum 
in (\ref{Ku}) denote the integer part, and $a_{nr}$ 
are all strictly positive. 
The sum term in (\ref{Ku}) is
$O\bigl(n^{-3}\bigr)$ as $n\to\infty$, and numerics shows that the sum
term contribution to $Q(n,z)$ 
is less than $1.1\%$ already 
for $n=1$ and less than $0.25\%$ for $n\ge2$. 

\begin{figure}[b!]
\includegraphics[scale=0.42]{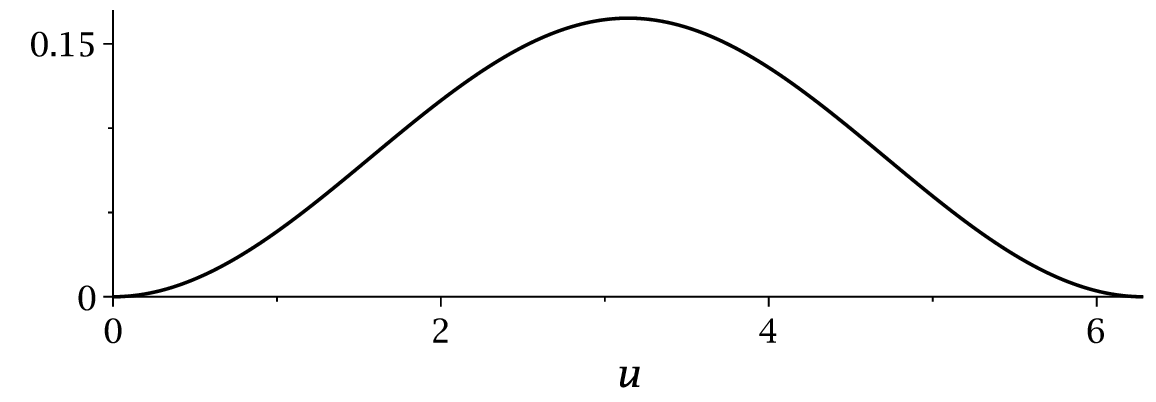}
\caption{\label{PNi}The plot shows $\left(\tfrac12 -
\Nboost\right)h^{-2}$
with $k=1$ 
as a function of
$u:= \tilde\Omega_1 \bar{\tau}$ over the full period
$0\le u \le2\pi$.} 
\end{figure}

$\Nboost$ (\ref{Hachi}) is periodic in $\bar\tau$
with period $2\pi\tilde\Omega_1^{-1}$ and attains its unique 
minimum at half-period. 
A~plot is shown in Figure~\ref{PNi}. 
The reason for the periodicity is that
the full time evolution of the field in 
Rob's cavity during the accelerated segment 
is periodic with this period 
since the frequencies $\tilde\Omega_n$
are integer multiples
of the fundamental frequency~$\tilde\Omega_1$.
$\Nboost$~is therefore periodic not just in the small $h$
approximation of (\ref{Hachi}) but exactly for arbitrary~$h$. 
More generally, the same 
periodicity occurs for all cavity trajectories that contain a uniformly
accelerated segment. 
We note that the period can be written as
$2\delta \, {(h/2)}^{-1}\atanh(h/2)$: this is the proper time
elapsed at the centre of the cavity between sending and recapturing a
light ray that bounces off each wall once.

\begin{figure}[b!]
\vspace*{-4ex}%
\includegraphics[scale=0.44]{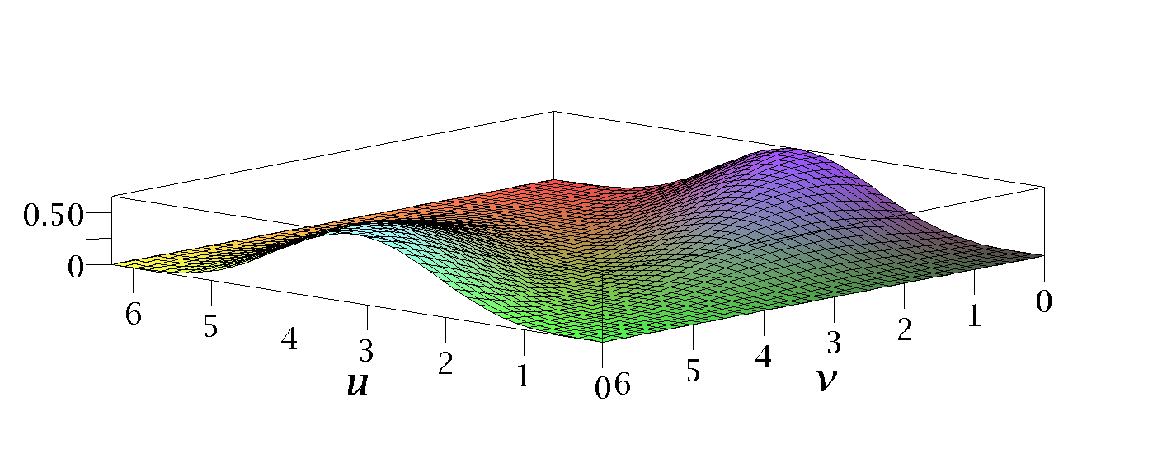}%
\vspace*{-2ex}%
\caption{\label{PSan}The plot shows 
$\left(\tfrac12 -
\Noneway\right)h^{-2}$
with $k=1$ as a function of
$u:= \tilde\Omega_1 \bar{\tau}$ and 
$v:= \pi\bar{\tau}'/\delta$
over the full periods
$0\le u \le2\pi$ and 
$0\le v \le2\pi$. Note the zeroes at $u\equiv0 \mod 2\pi$ and at 
$u+v \equiv 0 \mod 2\pi$.}
\end{figure}

As a second travel scenario, suppose that Rob blasts off as above,
coasts inertially for proper time ${\bar\tau}'$ and then performs a
braking manoeuvre that is the reverse of the initial acceleration,
bringing him to rest (at, say, Alpha Centauri).  Composing the
segments as above, we see that the negativity $\Noneway$ is periodic
in ${\bar\tau}'$ with period~$2\delta$. 
Noting that for leftward acceleration 
(\ref{eq:mrcoeffs}) holds with negative~$h$, 
we find to order $h^2$ the formula 
\begin{align}
\Noneway
& = 
\tfrac12 
- 
h^2 \sum_{r=0}^{\infty}
a_{kr} \, {\bigl|p^{1+2r} - 1\bigr|}^2 
{\bigl|{(pp')}^{1+2r} - 1\bigr|}^2  
\notag
\\[2ex]
& = \tfrac12 
- 2 \bigl[ 
2Q(k,1) -2Q(k,p) + Q(k,p^{\prime})
\nonumber\\[1ex]
& \hspace{10ex}
- 2Q(k,pp^{\prime}) 
+ Q(k,p^2p^{\prime})
\bigr] 
h^2 , 
\label{Ju}
\end{align}
where $p':= \exp\bigl(i\pi\bar{\tau}'/\delta\bigr)$. 
In addition to
displaying the periodicities in $\bar{\tau}$ and~$\bar{\tau}'$,
(\ref{Ju}) shows that the coefficient of $h^2$ vanishes
iff $p=1$ or $pp'=1$. This implies that to order~$h^2$, 
any entanglement degradation
caused by the accelerated segments can be cancelled by fine-tuning
the duration of the coasting segment. A~plot is shown in
Figure~\ref{PSan}. 

\begin{figure}
\includegraphics[scale=0.38]{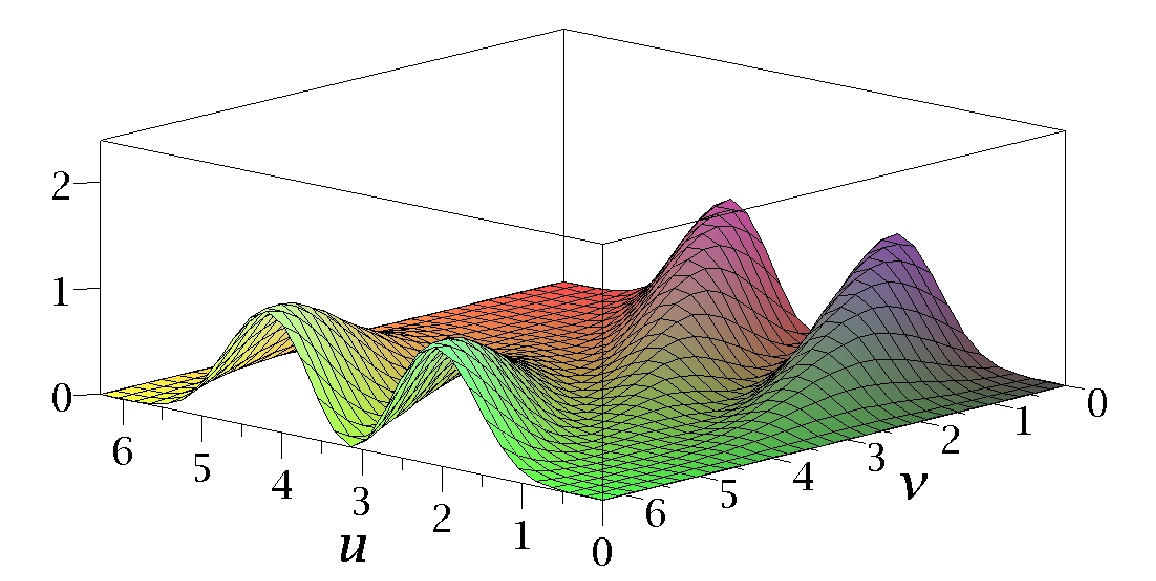}
\newline
$\phantom{xxx}$ 
\newline 
\includegraphics[scale=0.38]{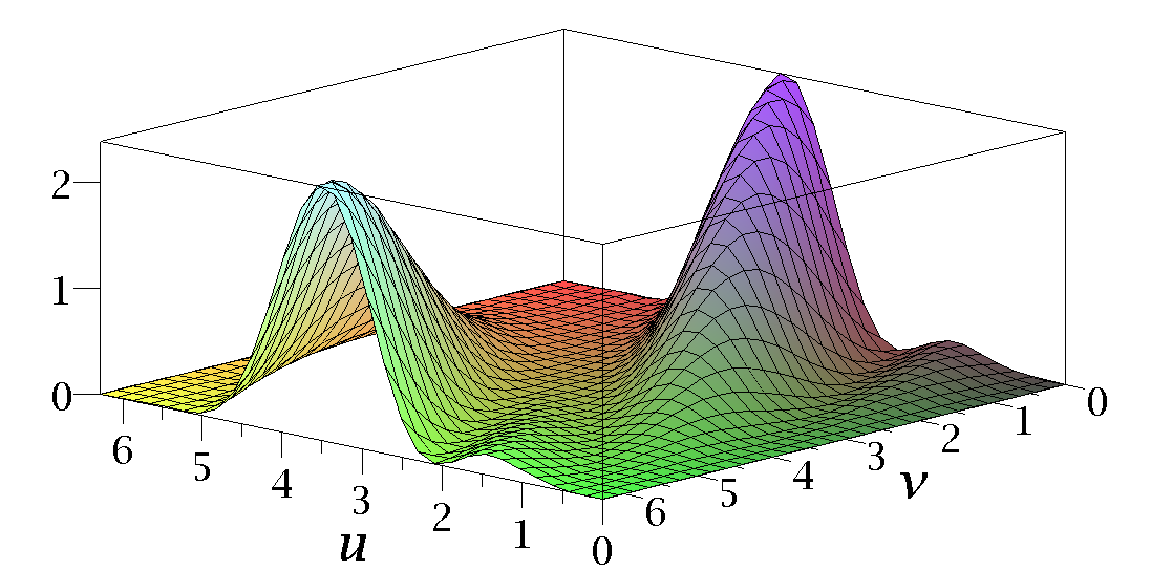}
\newline
$\phantom{xxx}$ 
\newline
\includegraphics[scale=0.38]{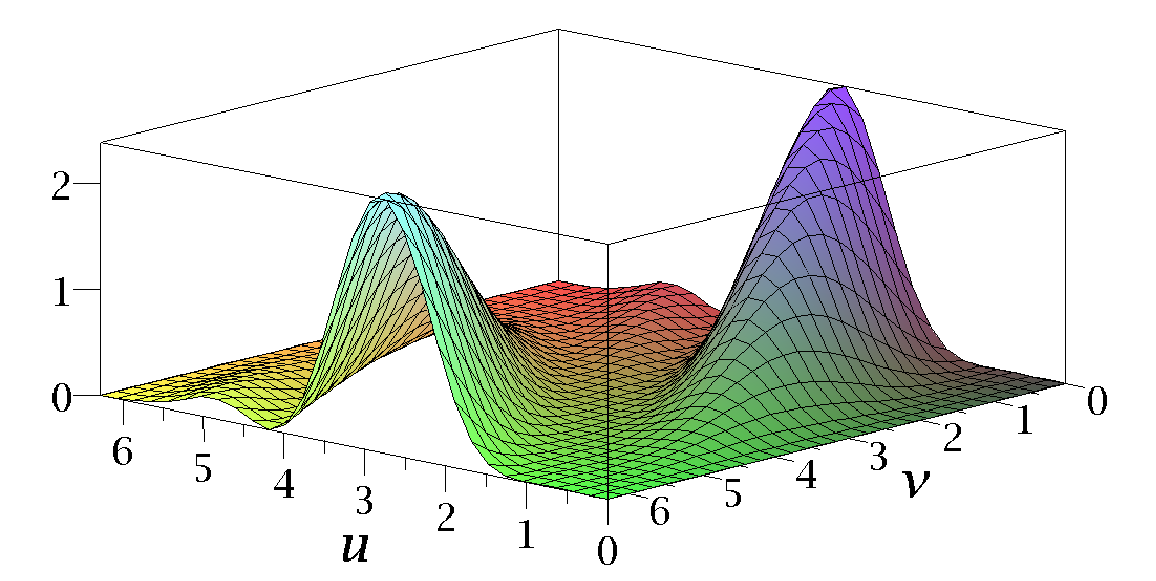}
\vspace*{1ex}
\caption{\label{fig:roundtrip}The plots show 
$\left(\tfrac12 -
\Ntwoway\right)h^{-2}$
with $k=1$ as a function of
$u:= \tilde\Omega_1 \bar{\tau}$ and 
$v:= \pi\bar{\tau}'/\delta$ 
for 
$0\le u \le2\pi$ and 
$0\le v \le2\pi$ when (from top to bottom) 
$\bar{\tau}'' = 0$, $2\delta/3$ and~$4\delta/3$.}
\end{figure}

As a third scenario, suppose Rob travels to Alpha Centauri as above,
rests there for proper time ${\bar\tau}''$ and then returns to Alice
by reversing the outward manoeuvres. Again composing the segments, we
see that the negativity $\Ntwoway$ is periodic in ${\bar\tau}''$ with
period~$2\delta$, and to order $h^2$ we find
\begin{align}
& \Ntwoway 
= 
\tfrac12 
- 
h^2 \sum_{r=0}^{\infty}
a_{kr} \, {\bigl|p^{1+2r} - 1\bigr|}^2 
{\bigl|{(pp')}^{1+2r} - 1\bigr|}^2 
\times 
\notag
\\
& \hspace{19.5ex}
\times 
{\bigl|{(p^2p'p'')}^{1+2r} - 1\bigr|}^2 , 
\label{eq:negroundtrip}
\end{align}
where $p'':= \exp\bigl(i\pi\bar{\tau}''/\delta\bigr)$, 
and the sum can be expressed as a sum of 14 $Q$s if desired 
[cf.~the final expressions in (\ref{Hachi}) and~(\ref{Ju})]. 
The
periodicites in $\bar{\tau}$, $\bar{\tau}'$ and $\bar{\tau}''$ are
manifest in~(\ref{eq:negroundtrip}). The coefficient of $h^2$ 
vanishes iff $p=1$, $pp'=1$ or
$p^2p'p''=1$, so that to order $h^2$ 
any entanglement degradation caused by the accelerated segments
can be cancelled by fine-tuning the duration of either of the 
independent inertial segments. 
Selected plots are shown in Figure~\ref{fig:roundtrip}.

Four comments are in order.  
First, should one wish to consider 
noninertial initial or final states for Rob, 
our small $h$ analysis is applicable whenever 
the assumptions leading to 
formula (\ref{Roku}) still hold. 
For example, in a kickstart scenario 
that contains just regions I and II of Figure~\ref{PIchi}, 
so that Rob's final state is uniformly accelerating, 
we find 
$\Nkick = \tfrac12 -Q(k,1) h^2$. 

Second, the
validity of our perturbative treatment requires the negativity to
remain close to its initial value~$\frac12$, which in our scenarios
happens when $kh\ll1$.  As the expansions (\ref{eq:mrcoeffs}) are not
uniform in the indices, the treatment could potentially have missed
even in this regime effects due to very high energy modes. However, we
have verified that when the $h^2$ terms are included in the
expansions~(\ref{eq:mrcoeffs}), these expansions satisfy the
Bogoliubov identities \cite{birrell-davies} perturbatively to order
$h^2$ and the products of the order $h$ matrices in the identities are
unconditionally convergent. This gives confidence in our order $h^2$
negativity formulas, whose infinite sums come from similar products of
order $h$ matrices.

Third, the matrices (\ref{eq:mrcoeffs}) 
can be self-consistently
truncated to the lowest $2\times2$ block provided the rows and columns
are renormalised by suitable factors of the form $1 + O(h^2)$ to
preserve the Bogoliubov identities to order~$h^2$.  Taking Rob's
initial excitation to be in the lower frequency, we find that all the
above negativity results hold with the replacement $Q(1,z) \to
a_{10}\Realpart(z) + \frac12 a_{11}\Realpart(z^3)$, and the error in
this replacement is less than~$0.7\%$. The high frequency effects on
the entanglement are hence very strongly suppressed. 

Fourth, the analysis can be adapted to a fermionic field 
\cite{PhysRevD.85.025012} and to 
scenarios where mode entanglement is generated from an initially
unentangled state~\cite{Friis:2012tb,*Bruschi:2012uf}.

\section{Massive field} 

\begin{figure}
\includegraphics[scale=0.4]{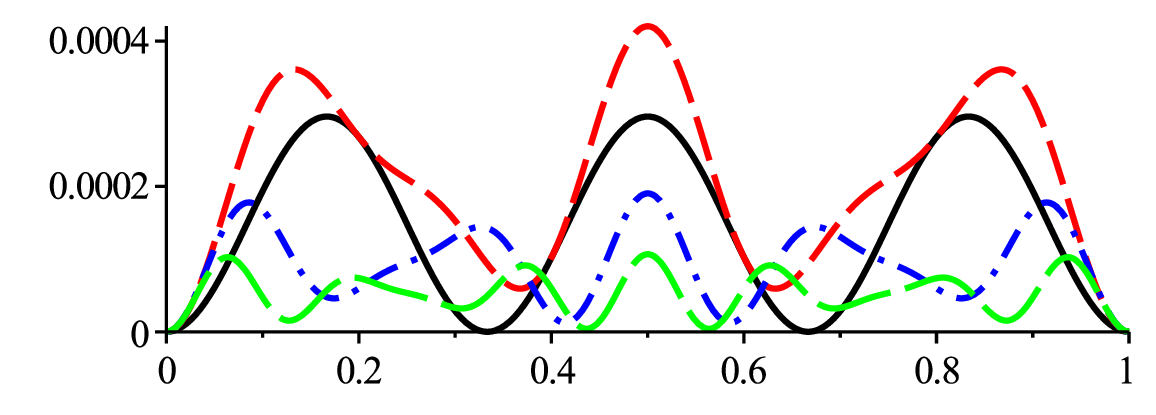}\\
\hspace*{-2ex}\includegraphics[scale=0.42]{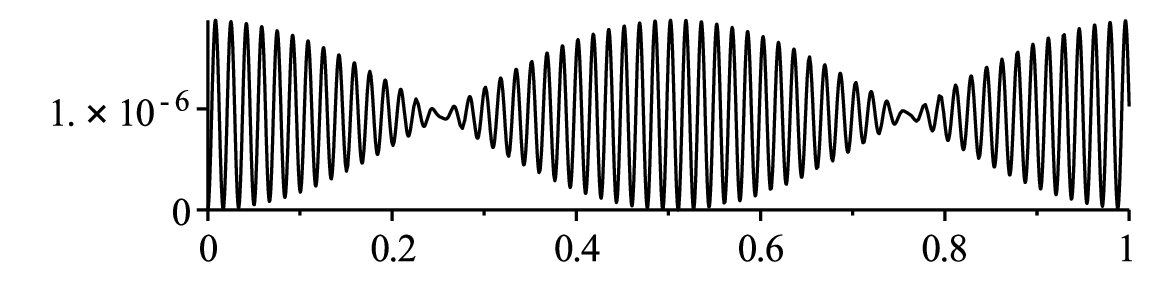}
\caption{\label{fig:massive}The plots show 
$\left(\tfrac12 -
\Nboost\right)h^{-2}M^{-4}$ (\ref{eq:massoneperiod}) 
for $M = 10^3$ 
as a function of 
$\pi\bar\tau/(4M\delta)$, in the upper figure with  
$k=1,2,3,4$ (solid, dashed, dash-dotted,
wide-dashed) and in the lower figure with $k=30$.}
\end{figure}

For a massive field the frequencies
are not uniformly spaced and the negativity is no longer periodic in
the durations of the inertial and uniformly accelerated segments. The
massive counterparts of the expansions (\ref{eq:mrcoeffs}) can be
found using uniform asymptotic expansions of modified Bessel functions
\cite{nist-dig-library,dunster:1990:bfp}, 
with the result 
\begin{widetext}
\begin{subequations}
\label{eq:MMRbogos}
\begin{align}
\mralpha_{nn} &= 1- 
\left(
{\frac {{\pi }^{2}{n}^{2}}{240}} 
+ {\frac {M^2}{120}} 
+ {\frac {M^2(M^2-5)}{240 \, \pi^2 n^2}} 
+ {\frac {M^2(M^2-24)}{96 \, \pi^4 n^4}}
- {\frac {7M^4}{16 \, \pi^6 n^6}}
\right) \! h^2 + O(h^4) , 
\hspace{2ex}
\mrbeta_{nn} = O(h^2) , 
\\
\mralpha_{mn} + \mrbeta_{mn}&= 
\frac{2 mn \bigl( -1+ {(-1)}^{m-n}
 \bigr) \bigl[\pi^2(n^2+3m^2) + 4M^2\bigr]
 {(M^2 + \pi^2 n^2)}^{1/4}}
{\pi^4 {(m^2-n^2)}^3 {(M^2 + \pi^2 m^2)}^{1/4}}
h 
+ O(h^2) 
\hspace{2ex}
(m\ne n), 
\\
\mralpha_{mn} - \mrbeta_{mn}&= 
\frac{2 mn \bigl( -1+ {(-1)}^{m-n}
 \bigr) \bigl[\pi^2(m^2+3n^2) + 4M^2\bigr]
 {(M^2 + \pi^2 m^2)}^{1/4}}
{\pi^4 {(m^2-n^2)}^3 {(M^2 + \pi^2 n^2)}^{1/4}}
h 
+ O(h^2) 
\hspace{2ex}
(m\ne n), 
\end{align} 
\end{subequations}
\end{widetext}
and we have again 
verified that the Bogoliubov
identities are  
satisfied perturbatively to order~$h^2$. 
The perturbative treatment is now valid
provided $h\ll1$ and $hM^2\lesssim100$, allowing the possibility
that $M$ may be large. 
When $k\ll M$, a 
qualitatively new feature is that the order $h$
contribution in $\mralpha$ is proportional to $M^2$,
resulting in an overall enhancement factor $M^4$ in the negativity. 
In the travel scenario with one accelerated segment, the negativity
takes in this limit the form 
\begin{align}
&\Nboost 
= \tfrac12 - h^2 M^4 \times \frac{256 k^2}{\pi^8}
\sideset{}{''}\sum_n 
\frac{n^2}{{(k^2 - n^2)}^6} \times 
\label{eq:massoneperiod}
\\
&
\hspace*{1ex}\times 
\left\{
1 - \cos\left[\bigl(\sqrt{M^2 + \pi^2 k^2} 
- \sqrt{M^2 + \pi^2 n^2}\,\bigr)(\bar\tau/\delta)\right]
\right\}, 
\notag
\end{align}
where the double prime means that the sum is 
over positive $n$ 
with $n \equiv k+1 \mod 2$. 
$\Nboost$~(\ref{eq:massoneperiod}) is approximately 
periodic in $\bar\tau$ 
with period 
$4M\delta/\pi$, 
but it contains also significant 
higher frequency components. 
Plots are shown in Figure~\ref{fig:massive}. 

\section{Discussion: $(3+1)$ dimensions} 

The above
$(1+1)$-dimensional entanglement degradation analysis extends
immediately to linear acceleration in $(3+1)$-dimensional Minkowski
space, where the transverse momentum merely contributes to the
effective $(1+1)$-dimensional mass. For a massless field in a cavity
of length $\delta = 10\,$m and acceleration $10\,$ms${}^{-2}$, an
effect of observable magnitude can be achieved by trapping quanta of
optical wavelengths provided the momentum is highly transverse so that
$k \ll M \approx 10^{8}$.  Were it possible to trap and stabilise
massive quanta of kaon mass $\mu = 10^{-27}\,$kg in a cavity of length
$\delta = 10\,$cm, the effect would become observable already at the
extreme microgravity acceleration of $10^{-10}\,$ms${}^{-2}$.

These estimates suggest that experimental verification of the effect
is feasible, but they also suggest that accelerations of gravitational
origin should be properly accounted for, both in the theoretical
analysis and in the experimental setup. On the theoretical side, we
anticipate that the core properties of our analysis extend to
nonlinear acceleration and to weakly curved spacetime.  To achieve
experimental control, especially on accelerations on gravitational
origin, the experiment may need to be performed by entangling cavities
in spaceships. An experiment within a single spaceship could test the
relative acceleration effect analysed in this paper. An experiment
using cavities in two and possibly widely separated spaceships would
test whether an equivalence principle between acceleration and a
gravitational field holds also in the context of quantum information.
An experimental confirmation that gravity degrades entanglement would
indeed provide a novel addition to the currently scarce experimental
evidence on quantum phenomena involving
gravity~\cite{Nesvizhevsky2002297,*Nesvizhevsky2011367}.  

\begin{acknowledgments}
We thank 
Gerardo Adesso, 
Andrzej Dragan, 
Daniele Faccio, 
Nicolai Friis, 
Lucia Hackerm\"uller, 
Antony Lee, 
Robert Mann 
and 
Eduardo Mart\'in-Mart\'inez 
for helpful discussions and correspondence. 
J.~L. thanks Frank Olver for correspondence on a misprint in formula
(4.15) in~\cite{dunster:1990:bfp}. 
I.~F. thanks EPSRC [CAF Grant EP/G00496X/2] for financial support. 
J.~L. was supported in part by STFC (UK). 

\end{acknowledgments}

\bibliography{boxtrips}

\begin{thebibliography}{33}%
\makeatletter
\providecommand \@ifxundefined [1]{%
 \@ifx{#1\undefined}
}%
\providecommand \@ifnum [1]{%
 \ifnum #1\expandafter \@firstoftwo
 \else \expandafter \@secondoftwo
 \fi
}%
\providecommand \@ifx [1]{%
 \ifx #1\expandafter \@firstoftwo
 \else \expandafter \@secondoftwo
 \fi
}%
\providecommand \natexlab [1]{#1}%
\providecommand \enquote  [1]{``#1''}%
\providecommand \bibnamefont  [1]{#1}%
\providecommand \bibfnamefont [1]{#1}%
\providecommand \citenamefont [1]{#1}%
\providecommand \href@noop [0]{\@secondoftwo}%
\providecommand \href [0]{\begingroup \@sanitize@url \@href}%
\providecommand \@href[1]{\@@startlink{#1}\@@href}%
\providecommand \@@href[1]{\endgroup#1\@@endlink}%
\providecommand \@sanitize@url [0]{\catcode `\\12\catcode `\$12\catcode
  `\&12\catcode `\#12\catcode `\^12\catcode `\_12\catcode `\%12\relax}%
\providecommand \@@startlink[1]{}%
\providecommand \@@endlink[0]{}%
\providecommand \url  [0]{\begingroup\@sanitize@url \@url }%
\providecommand \@url [1]{\endgroup\@href {#1}{\urlprefix }}%
\providecommand \urlprefix  [0]{URL }%
\providecommand \Eprint [0]{\href }%
\providecommand \doibase [0]{http://dx.doi.org/}%
\providecommand \selectlanguage [0]{\@gobble}%
\providecommand \bibinfo  [0]{\@secondoftwo}%
\providecommand \bibfield  [0]{\@secondoftwo}%
\providecommand \translation [1]{[#1]}%
\providecommand \BibitemOpen [0]{}%
\providecommand \bibitemStop [0]{}%
\providecommand \bibitemNoStop [0]{.\EOS\space}%
\providecommand \EOS [0]{\spacefactor3000\relax}%
\providecommand \BibitemShut  [1]{\csname bibitem#1\endcsname}%
\let\auto@bib@innerbib\@empty
\bibitem [{\citenamefont {Villoresi}\ \emph {et~al.}(2008)\citenamefont
  {Villoresi}, \citenamefont {Jennewein}, \citenamefont {Tamburini},
  \citenamefont {Aspelmeyer}, \citenamefont {Bonato}, \citenamefont {Ursin},
  \citenamefont {Pernechele}, \citenamefont {Luceri}, \citenamefont {Bianco},
  \citenamefont {Zeilinger},\ and\ \citenamefont
  {Barbieri}}]{Villoresi:2008p491}%
  \BibitemOpen
  \bibfield  {author} {\bibinfo {author} {\bibfnamefont {P.}~\bibnamefont
  {Villoresi}}, \bibinfo {author} {\bibfnamefont {T.}~\bibnamefont
  {Jennewein}}, \bibinfo {author} {\bibfnamefont {F.}~\bibnamefont
  {Tamburini}}, \bibinfo {author} {\bibfnamefont {M.}~\bibnamefont
  {Aspelmeyer}}, \bibinfo {author} {\bibfnamefont {C.}~\bibnamefont {Bonato}},
  \bibinfo {author} {\bibfnamefont {R.}~\bibnamefont {Ursin}}, \bibinfo
  {author} {\bibfnamefont {C.}~\bibnamefont {Pernechele}}, \bibinfo {author}
  {\bibfnamefont {V.}~\bibnamefont {Luceri}}, \bibinfo {author} {\bibfnamefont
  {G.}~\bibnamefont {Bianco}}, \bibinfo {author} {\bibfnamefont
  {A.}~\bibnamefont {Zeilinger}}, \ and\ \bibinfo {author} {\bibfnamefont
  {C.}~\bibnamefont {Barbieri}},\ }\href {\doibase
  10.1088/1367-2630/10/3/033038} {\bibfield  {journal} {\bibinfo  {journal}
  {New J. Phys.}\ }\textbf {\bibinfo {volume} {10}},\ \bibinfo {pages} {033038}
  (\bibinfo {year} {2008})}\BibitemShut {NoStop}%
\bibitem [{\citenamefont {Bonato}\ \emph {et~al.}(2009)\citenamefont {Bonato},
  \citenamefont {Tomaello}, \citenamefont {Deppo}, \citenamefont {Naletto},\
  and\ \citenamefont {Villoresi}}]{Bonato:2009p959}%
  \BibitemOpen
  \bibfield  {author} {\bibinfo {author} {\bibfnamefont {C.}~\bibnamefont
  {Bonato}}, \bibinfo {author} {\bibfnamefont {A.}~\bibnamefont {Tomaello}},
  \bibinfo {author} {\bibfnamefont {V.~D.}\ \bibnamefont {Deppo}}, \bibinfo
  {author} {\bibfnamefont {G.}~\bibnamefont {Naletto}}, \ and\ \bibinfo
  {author} {\bibfnamefont {P.}~\bibnamefont {Villoresi}},\ }\href {\doibase
  10.1088/1367-2630/11/4/045017} {\bibfield  {journal} {\bibinfo  {journal}
  {New J. Phys.}\ }\textbf {\bibinfo {volume} {11}},\ \bibinfo {pages} {045017}
  (\bibinfo {year} {2009})}\BibitemShut {NoStop}%
\bibitem [{\citenamefont {Downes}\ \emph {et~al.}(2011)\citenamefont {Downes},
  \citenamefont {Fuentes},\ and\ \citenamefont {Ralph}}]{Downes:2010tv}%
  \BibitemOpen
  \bibfield  {author} {\bibinfo {author} {\bibfnamefont {T.~G.}\ \bibnamefont
  {Downes}}, \bibinfo {author} {\bibfnamefont {I.}~\bibnamefont {Fuentes}}, \
  and\ \bibinfo {author} {\bibfnamefont {T.~C.}\ \bibnamefont {Ralph}},\ }\href
  {\doibase 10.1103/PhysRevLett.106.210502} {\bibfield  {journal} {\bibinfo
  {journal} {Phys. Rev. Lett.}\ }\textbf {\bibinfo {volume} {106}},\ \bibinfo
  {pages} {210502} (\bibinfo {year} {2011})}\BibitemShut {NoStop}%
\bibitem [{\citenamefont {Schlicht}(2004)}]{Schlicht:2003iy}%
  \BibitemOpen
  \bibfield  {author} {\bibinfo {author} {\bibfnamefont {S.}~\bibnamefont
  {Schlicht}},\ }\href {\doibase 10.1088/0264-9381/21/19/011} {\bibfield
  {journal} {\bibinfo  {journal} {Class. Quant. Grav.}\ }\textbf {\bibinfo
  {volume} {21}},\ \bibinfo {pages} {4647} (\bibinfo {year}
  {2004})}\BibitemShut {NoStop}%
\bibitem [{\citenamefont {Langlois}(2006)}]{Langlois:2005nf}%
  \BibitemOpen
  \bibfield  {author} {\bibinfo {author} {\bibfnamefont {P.}~\bibnamefont
  {Langlois}},\ }\href {\doibase 10.1016/j.aop.2006.01.013} {\bibfield
  {journal} {\bibinfo  {journal} {Annals Phys.}\ }\textbf {\bibinfo {volume}
  {321}},\ \bibinfo {pages} {2027} (\bibinfo {year} {2006})}\BibitemShut
  {NoStop}%
\bibitem [{\citenamefont {Louko}\ and\ \citenamefont
  {Satz}(2006)}]{Louko:2006zv}%
  \BibitemOpen
  \bibfield  {author} {\bibinfo {author} {\bibfnamefont {J.}~\bibnamefont
  {Louko}}\ and\ \bibinfo {author} {\bibfnamefont {A.}~\bibnamefont {Satz}},\
  }\href@noop {} {\bibfield  {journal} {\bibinfo  {journal} {Class. Quant.
  Grav.}\ }\textbf {\bibinfo {volume} {23}},\ \bibinfo {pages} {6321} (\bibinfo
  {year} {2006})}\BibitemShut {NoStop}%
\bibitem [{\citenamefont {Satz}(2007)}]{Satz:2006kb}%
  \BibitemOpen
  \bibfield  {author} {\bibinfo {author} {\bibfnamefont {A.}~\bibnamefont
  {Satz}},\ }\href {\doibase 10.1088/0264-9381/24/7/003} {\bibfield  {journal}
  {\bibinfo  {journal} {Class. Quant. Grav.}\ }\textbf {\bibinfo {volume}
  {24}},\ \bibinfo {pages} {1719} (\bibinfo {year} {2007})}\BibitemShut
  {NoStop}%
\bibitem [{\citenamefont {Louko}\ and\ \citenamefont
  {Satz}(2008)}]{Louko:2007mu}%
  \BibitemOpen
  \bibfield  {author} {\bibinfo {author} {\bibfnamefont {J.}~\bibnamefont
  {Louko}}\ and\ \bibinfo {author} {\bibfnamefont {A.}~\bibnamefont {Satz}},\
  }\href@noop {} {\bibfield  {journal} {\bibinfo  {journal} {Class. Quant.
  Grav.}\ }\textbf {\bibinfo {volume} {25}},\ \bibinfo {pages} {055012}
  (\bibinfo {year} {2008})}\BibitemShut {NoStop}%
\bibitem [{\citenamefont {Alsing}\ and\ \citenamefont
  {Milburn}(2003)}]{Alsingtelep}%
  \BibitemOpen
  \bibfield  {author} {\bibinfo {author} {\bibfnamefont {P.~M.}\ \bibnamefont
  {Alsing}}\ and\ \bibinfo {author} {\bibfnamefont {G.~J.}\ \bibnamefont
  {Milburn}},\ }\href {\doibase 10.1103/PhysRevLett.91.180404} {\bibfield
  {journal} {\bibinfo  {journal} {Phys. Rev. Lett.}\ }\textbf {\bibinfo
  {volume} {91}},\ \bibinfo {pages} {180404} (\bibinfo {year}
  {2003})}\BibitemShut {NoStop}%
\bibitem [{\citenamefont {Fuentes-Schuller}\ and\ \citenamefont
  {Mann}(2005)}]{FuentesSchuller:2005p47}%
  \BibitemOpen
  \bibfield  {author} {\bibinfo {author} {\bibfnamefont {I.}~\bibnamefont
  {Fuentes-Schuller}}\ and\ \bibinfo {author} {\bibfnamefont {R.~B.}\
  \bibnamefont {Mann}},\ }\href {\doibase 10.1103/PhysRevLett.95.120404}
  {\bibfield  {journal} {\bibinfo  {journal} {Phys. Rev. Lett.}\ }\textbf
  {\bibinfo {volume} {95}},\ \bibinfo {pages} {120404} (\bibinfo {year}
  {2005})}\BibitemShut {NoStop}%
\bibitem [{\citenamefont {Bruschi}\ \emph {et~al.}(2010)\citenamefont
  {Bruschi}, \citenamefont {Louko}, \citenamefont {Mart\'\i{}n-Mart\'\i{}nez},
  \citenamefont {Dragan},\ and\ \citenamefont {Fuentes}}]{PhysRevA.82.042332}%
  \BibitemOpen
  \bibfield  {author} {\bibinfo {author} {\bibfnamefont {D.~E.}\ \bibnamefont
  {Bruschi}}, \bibinfo {author} {\bibfnamefont {J.}~\bibnamefont {Louko}},
  \bibinfo {author} {\bibfnamefont {E.}~\bibnamefont
  {Mart\'\i{}n-Mart\'\i{}nez}}, \bibinfo {author} {\bibfnamefont
  {A.}~\bibnamefont {Dragan}}, \ and\ \bibinfo {author} {\bibfnamefont
  {I.}~\bibnamefont {Fuentes}},\ }\href {\doibase 10.1103/PhysRevA.82.042332}
  {\bibfield  {journal} {\bibinfo  {journal} {Phys. Rev. A}\ }\textbf {\bibinfo
  {volume} {82}},\ \bibinfo {pages} {042332} (\bibinfo {year}
  {2010})}\BibitemShut {NoStop}%
\bibitem [{\citenamefont {Lambrecht}\ \emph {et~al.}(1996)\citenamefont
  {Lambrecht}, \citenamefont {Jaekel},\ and\ \citenamefont
  {Reynaud}}]{PhysRevLett.77.615}%
  \BibitemOpen
  \bibfield  {author} {\bibinfo {author} {\bibfnamefont {A.}~\bibnamefont
  {Lambrecht}}, \bibinfo {author} {\bibfnamefont {M.-T.}\ \bibnamefont
  {Jaekel}}, \ and\ \bibinfo {author} {\bibfnamefont {S.}~\bibnamefont
  {Reynaud}},\ }\href {\doibase 10.1103/PhysRevLett.77.615} {\bibfield
  {journal} {\bibinfo  {journal} {Phys. Rev. Lett.}\ }\textbf {\bibinfo
  {volume} {77}},\ \bibinfo {pages} {615} (\bibinfo {year} {1996})}\BibitemShut
  {NoStop}%
\bibitem [{\citenamefont {Wilson}\ \emph {et~al.}(2011)\citenamefont {Wilson},
  \citenamefont {Johansson}, \citenamefont {Pourkabirian}, \citenamefont
  {Simoen}, \citenamefont {Johansson}, \citenamefont {Duty}, \citenamefont
  {Nori},\ and\ \citenamefont {Delsing}}]{wilson-casimir-obs}%
  \BibitemOpen
  \bibfield  {author} {\bibinfo {author} {\bibfnamefont {C.~M.}\ \bibnamefont
  {Wilson}}, \bibinfo {author} {\bibfnamefont {G.}~\bibnamefont {Johansson}},
  \bibinfo {author} {\bibfnamefont {A.}~\bibnamefont {Pourkabirian}}, \bibinfo
  {author} {\bibfnamefont {M.}~\bibnamefont {Simoen}}, \bibinfo {author}
  {\bibfnamefont {J.~R.}\ \bibnamefont {Johansson}}, \bibinfo {author}
  {\bibfnamefont {T.}~\bibnamefont {Duty}}, \bibinfo {author} {\bibfnamefont
  {F.}~\bibnamefont {Nori}}, \ and\ \bibinfo {author} {\bibfnamefont
  {P.}~\bibnamefont {Delsing}},\ }\href {\doibase 110.1038/nature10561}
  {\bibfield  {journal} {\bibinfo  {journal} {Nature}\ }\textbf {\bibinfo
  {volume} {479}},\ \bibinfo {pages} {376} (\bibinfo {year}
  {2011})}\BibitemShut {NoStop}%
\bibitem [{\citenamefont {Dodonov}(2010)}]{dodonovCasimirReview2010}%
  \BibitemOpen
  \bibfield  {author} {\bibinfo {author} {\bibfnamefont {V.~V.}\ \bibnamefont
  {Dodonov}},\ }\href {\doibase 10.1088/0031-8949/82/03/038105} {\bibfield
  {journal} {\bibinfo  {journal} {Phys. Scr.}\ }\textbf {\bibinfo {volume}
  {82}},\ \bibinfo {pages} {038105} (\bibinfo {year} {2010})}\BibitemShut
  {NoStop}%
\bibitem [{\citenamefont {Raimond}\ \emph {et~al.}(2001)\citenamefont
  {Raimond}, \citenamefont {Brune},\ and\ \citenamefont
  {Haroche}}]{Raimond:2001p6}%
  \BibitemOpen
  \bibfield  {author} {\bibinfo {author} {\bibfnamefont {J.~M.}\ \bibnamefont
  {Raimond}}, \bibinfo {author} {\bibfnamefont {M.}~\bibnamefont {Brune}}, \
  and\ \bibinfo {author} {\bibfnamefont {S.}~\bibnamefont {Haroche}},\ }\href
  {\doibase 10.1103/RevModPhys.73.565} {\bibfield  {journal} {\bibinfo
  {journal} {Rev. Mod. Phys.}\ }\textbf {\bibinfo {volume} {73}},\ \bibinfo
  {pages} {565} (\bibinfo {year} {2001})}\BibitemShut {NoStop}%
\bibitem [{\citenamefont {Browne}\ and\ \citenamefont
  {Plenio}(2003)}]{Browne:2003p374}%
  \BibitemOpen
  \bibfield  {author} {\bibinfo {author} {\bibfnamefont {D.~E.}\ \bibnamefont
  {Browne}}\ and\ \bibinfo {author} {\bibfnamefont {M.~B.}\ \bibnamefont
  {Plenio}},\ }\href {\doibase 10.1103/PhysRevA.67.012325} {\bibfield
  {journal} {\bibinfo  {journal} {Phys. Rev. A}\ }\textbf {\bibinfo {volume}
  {67}},\ \bibinfo {pages} {012325} (\bibinfo {year} {2003})}\BibitemShut
  {NoStop}%
\bibitem [{\citenamefont {Birrell}\ and\ \citenamefont
  {Davies}(1982)}]{birrell-davies}%
  \BibitemOpen
  \bibfield  {author} {\bibinfo {author} {\bibfnamefont {N.~D.}\ \bibnamefont
  {Birrell}}\ and\ \bibinfo {author} {\bibfnamefont {P.~C.~W.}\ \bibnamefont
  {Davies}},\ }\href@noop {} {\emph {\bibinfo {title} {Quantum Fields in Curved
  Space}}}\ (\bibinfo  {publisher} {Cambridge University Press, Cambridge,
  England},\ \bibinfo {year} {1982})\BibitemShut {NoStop}%
\bibitem [{\citenamefont {Fabbri}\ and\ \citenamefont
  {Navarro-Salas}(2005)}]{fabbri-navarro-salas}%
  \BibitemOpen
  \bibfield  {author} {\bibinfo {author} {\bibfnamefont {A.}~\bibnamefont
  {Fabbri}}\ and\ \bibinfo {author} {\bibfnamefont {J.}~\bibnamefont
  {Navarro-Salas}},\ }\href@noop {} {\emph {\bibinfo {title} {Modeling Black
  Hole Evaporation}}}\ (\bibinfo  {publisher} {Imperial College Press, London,
  England},\ \bibinfo {year} {2005})\BibitemShut {NoStop}%
\bibitem [{\citenamefont {Vidal}\ and\ \citenamefont
  {Werner}(2002)}]{vidal-werner:computable}%
  \BibitemOpen
  \bibfield  {author} {\bibinfo {author} {\bibfnamefont {G.}~\bibnamefont
  {Vidal}}\ and\ \bibinfo {author} {\bibfnamefont {R.~F.}\ \bibnamefont
  {Werner}},\ }\href {\doibase 10.1103/PhysRevA.65.032314} {\bibfield
  {journal} {\bibinfo  {journal} {Phys. Rev. A}\ }\textbf {\bibinfo {volume}
  {65}},\ \bibinfo {pages} {032314} (\bibinfo {year} {2002})}\BibitemShut
  {NoStop}%
\bibitem [{\citenamefont {Audenaert}\ \emph {et~al.}(2003)\citenamefont
  {Audenaert}, \citenamefont {Plenio},\ and\ \citenamefont
  {Eisert}}]{audenart-plenio-eisert:cost}%
  \BibitemOpen
  \bibfield  {author} {\bibinfo {author} {\bibfnamefont {K.}~\bibnamefont
  {Audenaert}}, \bibinfo {author} {\bibfnamefont {M.~B.}\ \bibnamefont
  {Plenio}}, \ and\ \bibinfo {author} {\bibfnamefont {J.}~\bibnamefont
  {Eisert}},\ }\href {\doibase 10.1103/PhysRevLett.90.027901} {\bibfield
  {journal} {\bibinfo  {journal} {Phys. Rev. Lett.}\ }\textbf {\bibinfo
  {volume} {90}},\ \bibinfo {pages} {027901} (\bibinfo {year}
  {2003})}\BibitemShut {NoStop}%
\bibitem [{\citenamefont {Plenio}\ and\ \citenamefont
  {Virmani}(2007)}]{plenio-virmani:review}%
  \BibitemOpen
  \bibfield  {author} {\bibinfo {author} {\bibfnamefont {M.~B.}\ \bibnamefont
  {Plenio}}\ and\ \bibinfo {author} {\bibfnamefont {S.}~\bibnamefont
  {Virmani}},\ }\href@noop {} {\bibfield  {journal} {\bibinfo  {journal}
  {Quant. Inf. Comput.}\ }\textbf {\bibinfo {volume} {7}},\ \bibinfo {pages}
  {1} (\bibinfo {year} {2007})}\BibitemShut {NoStop}%
\bibitem [{\citenamefont {Roghani}\ \emph {et~al.}(2011)\citenamefont
  {Roghani}, \citenamefont {Helm},\ and\ \citenamefont
  {Breuer}}]{PhysRevLett.106.040502}%
  \BibitemOpen
  \bibfield  {author} {\bibinfo {author} {\bibfnamefont {M.}~\bibnamefont
  {Roghani}}, \bibinfo {author} {\bibfnamefont {H.}~\bibnamefont {Helm}}, \
  and\ \bibinfo {author} {\bibfnamefont {H.-P.}\ \bibnamefont {Breuer}},\
  }\href {\doibase 10.1103/PhysRevLett.106.040502} {\bibfield  {journal}
  {\bibinfo  {journal} {Phys. Rev. Lett.}\ }\textbf {\bibinfo {volume} {106}},\
  \bibinfo {pages} {040502} (\bibinfo {year} {2011})}\BibitemShut {NoStop}%
\bibitem [{\citenamefont {Gauger}\ \emph {et~al.}(2011)\citenamefont {Gauger},
  \citenamefont {Rieper}, \citenamefont {Morton}, \citenamefont {Benjamin},\
  and\ \citenamefont {Vedral}}]{PhysRevLett.106.040503}%
  \BibitemOpen
  \bibfield  {author} {\bibinfo {author} {\bibfnamefont {E.~M.}\ \bibnamefont
  {Gauger}}, \bibinfo {author} {\bibfnamefont {E.}~\bibnamefont {Rieper}},
  \bibinfo {author} {\bibfnamefont {J.~J.~L.}\ \bibnamefont {Morton}}, \bibinfo
  {author} {\bibfnamefont {S.~C.}\ \bibnamefont {Benjamin}}, \ and\ \bibinfo
  {author} {\bibfnamefont {V.}~\bibnamefont {Vedral}},\ }\href {\doibase
  10.1103/PhysRevLett.106.040503} {\bibfield  {journal} {\bibinfo  {journal}
  {Phys. Rev. Lett.}\ }\textbf {\bibinfo {volume} {106}},\ \bibinfo {pages}
  {040503} (\bibinfo {year} {2011})}\BibitemShut {NoStop}%
\bibitem [{\citenamefont {Di~Franco}\ \emph {et~al.}(2011)\citenamefont
  {Di~Franco}, \citenamefont {Mc~Gettrick},\ and\ \citenamefont
  {Busch}}]{PhysRevLett.106.080502}%
  \BibitemOpen
  \bibfield  {author} {\bibinfo {author} {\bibfnamefont {C.}~\bibnamefont
  {Di~Franco}}, \bibinfo {author} {\bibfnamefont {M.}~\bibnamefont
  {Mc~Gettrick}}, \ and\ \bibinfo {author} {\bibfnamefont {T.}~\bibnamefont
  {Busch}},\ }\href {\doibase 10.1103/PhysRevLett.106.080502} {\bibfield
  {journal} {\bibinfo  {journal} {Phys. Rev. Lett.}\ }\textbf {\bibinfo
  {volume} {106}},\ \bibinfo {pages} {080502} (\bibinfo {year}
  {2011})}\BibitemShut {NoStop}%
\bibitem [{\citenamefont {Fr\"owis}\ and\ \citenamefont
  {D\"ur}(2011)}]{PhysRevLett.106.110402}%
  \BibitemOpen
  \bibfield  {author} {\bibinfo {author} {\bibfnamefont {F.}~\bibnamefont
  {Fr\"owis}}\ and\ \bibinfo {author} {\bibfnamefont {W.}~\bibnamefont
  {D\"ur}},\ }\href {\doibase 10.1103/PhysRevLett.106.110402} {\bibfield
  {journal} {\bibinfo  {journal} {Phys. Rev. Lett.}\ }\textbf {\bibinfo
  {volume} {106}},\ \bibinfo {pages} {110402} (\bibinfo {year}
  {2011})}\BibitemShut {NoStop}%
\bibitem [{\citenamefont {Bar-Gill}\ \emph {et~al.}(2011)\citenamefont
  {Bar-Gill}, \citenamefont {Gross}, \citenamefont {Mazets}, \citenamefont
  {Oberthaler},\ and\ \citenamefont {Kurizki}}]{PhysRevLett.106.120404}%
  \BibitemOpen
  \bibfield  {author} {\bibinfo {author} {\bibfnamefont {N.}~\bibnamefont
  {Bar-Gill}}, \bibinfo {author} {\bibfnamefont {C.}~\bibnamefont {Gross}},
  \bibinfo {author} {\bibfnamefont {I.}~\bibnamefont {Mazets}}, \bibinfo
  {author} {\bibfnamefont {M.}~\bibnamefont {Oberthaler}}, \ and\ \bibinfo
  {author} {\bibfnamefont {G.}~\bibnamefont {Kurizki}},\ }\href {\doibase
  10.1103/PhysRevLett.106.120404} {\bibfield  {journal} {\bibinfo  {journal}
  {Phys. Rev. Lett.}\ }\textbf {\bibinfo {volume} {106}},\ \bibinfo {pages}
  {120404} (\bibinfo {year} {2011})}\BibitemShut {NoStop}%
\bibitem [{nis(2010)}]{nist-dig-library}%
  \BibitemOpen
  \href {http://dlmf.nist.gov/} {\emph {\bibinfo {title} {Digital Library of
  Mathematical Functions}}}\ (\bibinfo  {publisher} {National Institute of
  Standards and Technology},\ \bibinfo {year} {2010})\BibitemShut {NoStop}%
\bibitem [{\citenamefont {Friis}\ \emph
  {et~al.}(2012{\natexlab{a}})\citenamefont {Friis}, \citenamefont {Lee},
  \citenamefont {Bruschi},\ and\ \citenamefont {Louko}}]{PhysRevD.85.025012}%
  \BibitemOpen
  \bibfield  {author} {\bibinfo {author} {\bibfnamefont {N.}~\bibnamefont
  {Friis}}, \bibinfo {author} {\bibfnamefont {A.~R.}\ \bibnamefont {Lee}},
  \bibinfo {author} {\bibfnamefont {D.~E.}\ \bibnamefont {Bruschi}}, \ and\
  \bibinfo {author} {\bibfnamefont {J.}~\bibnamefont {Louko}},\ }\href
  {\doibase 10.1103/PhysRevD.85.025012} {\bibfield  {journal} {\bibinfo
  {journal} {Phys. Rev. D}\ }\textbf {\bibinfo {volume} {85}},\ \bibinfo
  {pages} {025012} (\bibinfo {year} {2012}{\natexlab{a}})}\BibitemShut
  {NoStop}%
\bibitem [{\citenamefont {Friis}\ \emph
  {et~al.}(2012{\natexlab{b}})\citenamefont {Friis}, \citenamefont {Bruschi},
  \citenamefont {Louko},\ and\ \citenamefont {Fuentes}}]{Friis:2012tb}%
  \BibitemOpen
  \bibfield  {author} {\bibinfo {author} {\bibfnamefont {N.}~\bibnamefont
  {Friis}}, \bibinfo {author} {\bibfnamefont {D.~E.}\ \bibnamefont {Bruschi}},
  \bibinfo {author} {\bibfnamefont {J.}~\bibnamefont {Louko}}, \ and\ \bibinfo
  {author} {\bibfnamefont {I.}~\bibnamefont {Fuentes}},\ }\href@noop {} {\
  (\bibinfo {year} {2012}{\natexlab{b}})},\ \Eprint
  {http://arxiv.org/abs/1201.0549} {arXiv:1201.0549 [quant-ph]} \BibitemShut
  {NoStop}%
\bibitem [{\citenamefont {Bruschi}\ \emph {et~al.}(2012)\citenamefont
  {Bruschi}, \citenamefont {Dragan}, \citenamefont {Lee}, \citenamefont
  {Fuentes},\ and\ \citenamefont {Louko}}]{Bruschi:2012uf}%
  \BibitemOpen
  \bibfield  {author} {\bibinfo {author} {\bibfnamefont {D.~E.}\ \bibnamefont
  {Bruschi}}, \bibinfo {author} {\bibfnamefont {A.}~\bibnamefont {Dragan}},
  \bibinfo {author} {\bibfnamefont {A.~R.}\ \bibnamefont {Lee}}, \bibinfo
  {author} {\bibfnamefont {I.}~\bibnamefont {Fuentes}}, \ and\ \bibinfo
  {author} {\bibfnamefont {J.}~\bibnamefont {Louko}},\ }\href@noop {} {\
  (\bibinfo {year} {2012})},\ \Eprint {http://arxiv.org/abs/1201.0663}
  {arXiv:1201.0663 [quant-ph]} \BibitemShut {NoStop}%
\bibitem [{\citenamefont {Dunster}(1990)}]{dunster:1990:bfp}%
  \BibitemOpen
  \bibfield  {author} {\bibinfo {author} {\bibfnamefont {T.~M.}\ \bibnamefont
  {Dunster}},\ }\href@noop {} {\bibfield  {journal} {\bibinfo  {journal} {SIAM
  J. Math. Anal.}\ }\textbf {\bibinfo {volume} {21}},\ \bibinfo {pages} {995}
  (\bibinfo {year} {1990})}\BibitemShut {NoStop}%
\bibitem [{\citenamefont {Nesvizhevsky}\ \emph {et~al.}(2002)\citenamefont
  {Nesvizhevsky}, \citenamefont {B\"{o}rner}, \citenamefont {Petukhov},
  \citenamefont {Abele}, \citenamefont {Bae\ss{}ler}, \citenamefont {Rue\ss},
  \citenamefont {St\"{o}ferle}, \citenamefont {Westphal}, \citenamefont
  {Gagarski}, \citenamefont {Guennady},\ and\ \citenamefont
  {Strelkov}}]{Nesvizhevsky2002297}%
  \BibitemOpen
  \bibfield  {author} {\bibinfo {author} {\bibfnamefont {V.~V.}\ \bibnamefont
  {Nesvizhevsky}}, \bibinfo {author} {\bibfnamefont {H.~G.}\ \bibnamefont
  {B\"{o}rner}}, \bibinfo {author} {\bibfnamefont {A.~K.}\ \bibnamefont
  {Petukhov}}, \bibinfo {author} {\bibfnamefont {H.}~\bibnamefont {Abele}},
  \bibinfo {author} {\bibfnamefont {S.}~\bibnamefont {Bae\ss{}ler}}, \bibinfo
  {author} {\bibfnamefont {F.~J.}\ \bibnamefont {Rue\ss}}, \bibinfo {author}
  {\bibfnamefont {T.}~\bibnamefont {St\"{o}ferle}}, \bibinfo {author}
  {\bibfnamefont {A.}~\bibnamefont {Westphal}}, \bibinfo {author}
  {\bibfnamefont {A.~M.}\ \bibnamefont {Gagarski}}, \bibinfo {author}
  {\bibfnamefont {A.}~\bibnamefont {Guennady}}, \ and\ \bibinfo {author}
  {\bibfnamefont {V.}~\bibnamefont {Strelkov}},\ }\href@noop {} {\bibfield
  {journal} {\bibinfo  {journal} {Nature}\ }\textbf {\bibinfo {volume} {415}},\
  \bibinfo {pages} {297} (\bibinfo {year} {2002})}\BibitemShut {NoStop}%
\bibitem [{\citenamefont {Nesvizhevsky}(2011)}]{Nesvizhevsky2011367}%
  \BibitemOpen
  \bibfield  {author} {\bibinfo {author} {\bibfnamefont {V.~V.}\ \bibnamefont
  {Nesvizhevsky}},\ }\href@noop {} {\bibfield  {journal} {\bibinfo  {journal}
  {Low Temperat. Phys.}\ }\textbf {\bibinfo {volume} {37}},\ \bibinfo {pages}
  {367} (\bibinfo {year} {2011})}\BibitemShut {NoStop}%
\end{thebibliography}%

\end{document}